%% file: samplepaper.tex
\documentclass[runningheads]{llncs}
\usepackage[T1]{fontenc}
\usepackage{graphicx}
\usepackage{booktabs}

\usepackage{xspace}
\newcommand{\name}{DPM-Bench\xspace}
\begin{document}
\title{DPM-Bench: Benchmark for Distributed Process Mining Algorithms on Cyber-Physical Systems}

\author{%
Hendrik Reiter\orcidID{0009-0003-8544-0012} 
\and Patrick Rathje\orcidID{0000-0003-3718-7115} 
\and Olaf Landsiedel\orcidID{0000-0001-6432-300X} 
\and Wilhelm Hasselbring\orcidID{0000-0001-6625-4335} 
}
\authorrunning{H. Reiter et al.}
\titlerunning{Benchmark for Distributed Process Mining}
%
\institute{Kiel University, Christian-Albrechts-Platz 4, 24118 Kiel, Germany \\ 
\email{\{hendrik.reiter, patrick.rathje, olaf.landsiedel, hasselbring\}@email.uni-kiel.de}}

\maketitle              
\begin{abstract}
Process Mining is established in research and industry systems to analyze and optimize processes based on event data from information systems. Within this work, we accomodate process mining techniques to Cyber-Physical Systems (CPS). To capture the distributed and heterogeneous characteristics of data, computational resources, and network communication in CPS, the today's process mining algorithms and techniques must be augmented. Specifically, there is a need for new \textit{Distributed} Process Mining algorithms that enable computations to be performed directly on edge resources, eliminating the need for moving all data to central cloud systems. This paper introduces the DPM-Bench benchmark for comparing such Distributed Process Mining algorithms. DPM-Bench is used to compare algorithms deployed in different computational topologies. The results enable information system engineers to assess whether the existing infrastructure is sufficient to perform distributed process mining, or to identify required improvements in algorithms and hardware. We present and discuss an experimental evaluation with DPM-Bench.

\end{abstract}

\input{paper}
\end{document}

%% file: paper.tex
\section{Introduction}

Cyber-Physical Systems (CPS)~\cite{baheti}  encapsulate the interplay between computational resources, networks, and physical components. Driven by the advancements of Industry 4.0, CPS have gained significant traction, particularly within the context of Smart Factories. The objective is to achieve fully automated and customizable production processes, which can be realized by integrating robots and other manufacturing equipment with computational capabilities. To facilitate human understanding and optimization of CPS, it is imperative to comprehend the underlying processes. Process mining~\cite{vanderAalst2022} is a discipline that enables the reconstruction, evaluation, prediction, and improvement of processes based on data extracted from information systems.

When an information systems engineer seeks to extract process data from production machinery, they encounter the challenge of efficiently utilizing available computing resources. Appliances within a factory are interconnected via a network and can execute computational tasks locally. For more complex calculations, additional computational capacities are leveraged, often found in cloud computing environments. This combination of local computational resources and servers in external data centers is commonly referred to as the (edge) cloud continuum~\cite{Moreschini2022}. However, offloading data to the cloud frequently presents several drawbacks. For instance, transmitting large datasets over the network is necessary, which can be time-consuming and resource-intensive. Additionally, data privacy~\cite{Ometov2022} and GDPR-conformance~\cite{Russo2018} must be addressed. Furthermore, each additional allocated computing resource incurs increased energy consumption~\cite{Khan2021} and associated costs. Consequently, performing as many computations as possible locally on edge devices is advantageous.

This is where our research starts. The local computational capacities in production appliances are often limited. They have limited computing power, memory capacity, and network bandwidth. In contrast to the distributed nature of event data, Process Mining traditionally assumes a central event log. If data cannot be shared due to data privacy reasons or scalability concerns, this requires a new class of process mining algorithms: \textit{Distributed Process Mining} (DPM).
DPM still lacks certain foundational elements in the research landscape. While initial attempts at DPM algorithms exist, there is a scarcity of formalisms to model the distribution inherent in both data and computations. Moreover, a comprehensive benchmark for evaluating these algorithms, considering the characteristics of computational nodes and network topologies, is absent. Such a benchmark is essential to determine the conditions under which DPM is beneficial and the trade-offs involved.

Our work contributes as follows:
\begin{enumerate}
    \item We motivate the opportunities of Distributed Process Mining, which processes the data near its sources
    \item We propose a taxonomy of topologies for Process Mining
    \item We provide a formalism to express the distribution of data and computing resources. The formalism supports the computing-centric quality measures.
    \item We developed \name to evaluate and compare prototypes of newly invented Distributed Process Mining algorithms.
\end{enumerate}

This work does not aim to simulate the technical details of CPS in exhaustive detail. Furthermore, it does not try to sell either Distributed Process Mining or a specific DPM algorithm. Instead, we provide a benchmark that enables a simple comparison of qualitative differences in algorithm and topology design.

The remainder of this paper is structured as follows: Section~2 motivates the application of Distributed Process Mining using a case study of a Smart Factory. Section~3 discusses related work.
Further, Section~4 establishes the fundamentals of Process Mining and Cyber-Physical Systems. Section~5 presents our formal foundations for DPM and formalizes distributed event logs and the computing nodes. Section~6 introduces \name and derives computing related quality measures.
Section~7 evaluates our benchmark by comparing two examples of Distributed Process Discovery algorithms in different computing topologies. Section~8 discusses the current state of \name, while Section~9 concludes the paper.

\section{Motivating Example}

\begin{figure}
    \centering
    \includegraphics[width=\textwidth]{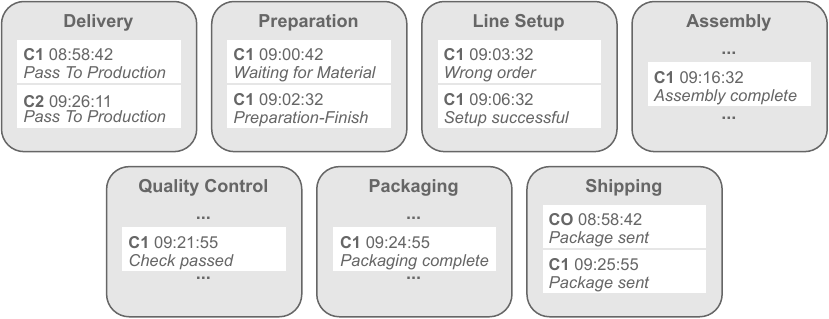}
    \caption{Exemplary Distributed Event Log for Processes in a Smart Factory: Events are distributed across different departments and are hence localized.}
    \label{fig:dlog}
\end{figure}

In a smart factory, various appliances and workstations are equipped with sensors and small computing nodes. For the purposes of this paper, we will use the example of a Smart Factory with a total of seven stations, as described in~\cite{reiter2024}.
A manufacturing process is modelled, involving goods delivery, material preparation, assembly line setup, assembling quality control, packaging, and shipping. The process is subject to potential waiting times and error-induced terminations. An example event log is presented in Figure~\ref{fig:dlog}. In contrast to standard process mining, this scenario lacks a centralized event log. Instead, each machine generates its own local event log. Moreover, each station has a local computing node for data collection and processing, enabling process discovery at its source. The advantages of this approach are exemplified by a quality control camera. Data transmission is minimized by performing local pattern recognition and transmitting only the derived activities. Furthermore, privacy-sensitive data, such as unintentionally recorded employee behavior, remains confined to the device. 
Process mining can provide real-time insights into processes and their variations within this scenario, enhancing their comprehensibility. An information systems engineer who intends to employ such a system has to evaluate the suitability of existing hardware in a factory. Two key questions arise: 1. Given the current hardware, what is the maximum data load that can be processed? 2. For the anticipated data load, how much additional hardware is required to supplement the existing infrastructure? Determining these metrics is a prerequisite for evaluating the feasibility and associated costs of implementing Distributed Process Mining.

\section{Related Work}

Related works encompass distribution in process mining, process mining in the realm of IoT, and edge-cloud simulations. Data distribution is already addressed in Federated Process Mining~\cite{FederatedPM,FederatedPM2}, which focuses on the level of the information system between multiple organizations, particularly on privacy protection. Several works address the distribution of calculations for process mining using big data frameworks such as MapReduce~\cite{MapReduce,FHM} to accelerate computations. Computational requirements are imposed on process mining algorithms, especially in Stream Process Mining~\cite{Burattin2022}, where real-time data processing requires specific runtime and memory complexity requirements. There are also initial algorithms that perform Distributed Process Mining at the edge level, as demonstrated in~\cite{Andersen}. Process mining in IoT has already been covered in \cite{Bertrand2022}. The potentials and applications for process mining in the Industry 4.0 are mentioned in~\cite{Osman2019,Vila2023}. The user and privacy challenges in IoT scenarios are discussed in~\cite{Michael2019}. However, they focus more on the context and data perspective and do not cover the computing perspective. For the latter, this paper introduces \name but other simulators for edge-cloud environments exist, like~\cite{Mass2020,Jha2020,Mechalikh2021}. However, those simulators have not been used in the context of process mining.

\section{Formal Foundation}

This section establishes the base concepts for this work. The first part delves into the formal underpinnings of Process Mining. Specifically, we explore process discovery by extracting a process model from an event log. In addition, the concept of real-time processing via event streams is introduced. The second part lays the groundwork for cyber-physical systems and edge computing.

\subsection{Process Mining}

Process mining is a discipline that focuses on the discovery, analysis, improvement, evaluation, simulation, and prediction of processes based on historical data~\cite{vanderAalst2022}. The subdiscipline \textit{Process discovery} involves creating a process model from an event log. An event represents the occurrence of an activity at a specific time associated with a particular case. Optionally, an additional payload can be attached to an event. An event log is an ordered, finite list of events. A process model is a representation of the process in an (ideally) human-understandable form. A common process model is the Directly-Follows-Graph (DFG), which visualizes the frequency with which two activities directly follow each other within a case.

In online or stream process mining~\cite{Burattin2022}, instead of a static event log, there is a continuously generated event stream without a predefined end. Stream process mining algorithms must operate within strict time and memory constraints (preferably constant runtime and space complexity), enabling real-time analysis~\cite{Burattin2022}. Moreover, the algorithm's results should be available at any given time.
\begin{definition}[Event and Event Stream]
An Event $e = [c, a, t, l]$ is a tuple of case $c \in \mathcal{C}$, activity $a \in \mathcal{A}$, timestamp $t \in \mathcal{T}$ and a location $l \in \mathcal{L}$. Let $\mathcal{E}$ denote  the set of all possible events defined as $\mathcal{E} := \mathcal{C}\times \mathcal{A}\times \mathcal{T}\times \mathcal{L}.$
An Event Stream $S$ is a function from the natural numbers to an event $e \in \mathcal{E}$, i.e., $S:= \mathbb{N} \rightarrow \mathcal{E}$. $S(i)$ indicates the i-th element of the event stream.
\end{definition}
The results of the stream discovery algorithm are retrieved using a process model request. Within this paper, we use an instance of the Directly-Follows-Graph (DFG) as the Process Model. According to van der Aalst, this DFG is a counted set of two directly following activities~\cite{Foundation}. 

\subsection{Cyber-Physical Systems}

\begin{figure}[H]
    \centering
    \includegraphics[width=.85\textwidth]{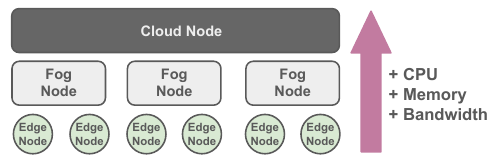}
    \caption{Data flow and computing capabilities in the Edge-Cloud-Continuum}
    \label{fig:ecc}
\end{figure}

Cyber-physical systems are defined as systems that possess physical components and are equipped with integrated computational resources~\cite{baheti}. These systems can also interact with humans, a concept referred to as human-cyber-physical systems~\cite{Wang2022}. The physical components generate data via integrated sensors ~\cite{Lee2015}. This data is produced continuously, in large volumes, and at high velocity, resulting in the traditional challenges associated with big data. Actuators can intervene in the processes within CPS as they control and optimize the system.
Wireless connections between the edge computing nodes allow them to communicate and, therefore, aggregate and share intermediate results~\cite{AHMADI2018}. This collaboration forms a distributed system of resource-limited devices. Edge computing~\cite{Cao2020} refers to the practice of processing algorithms near their data source. This contrasts cloud computing, which leverages remote, high-performance servers. Frequently, a hybrid approach combining edge and cloud resources is adopted.
Computational resources within CPS are often structured as a tree \cite{Bonomi2012,Hong2013}. 
Interconnected edge nodes with limited computational power, memory capacity, and network bandwidth form the foundation. Due to their close geographical proximity, communication between nodes can be achieved with low latency.
Above this, local fog nodes and global cloud resources are allocated.
In contrast to edge nodes, cloud nodes have stronger computing capabilities, more memory, and are connected via a network with higher bandwidth. Their downside is a higher network latency.
Figure~\ref{fig:ecc} illustrates the data flow and growing computing capabilities in the Edge-Cloud-Continuum \cite{Moreschini2022}.

\section{Distributed Process Mining}
This section provides a formal description of \name. We begin by classifying the topologies for Process Mining. Following this, we identify the shortcomings of existing stream process mining models in the context of cyber-physical systems. Based on the identified limitations, we extend the formalism. 

\subsection{Taxonomy of Process Mining Topologies}

The placement of the computing nodes is relevant when performing process mining on the aforementioned production line example. We propose a taxonomy of topologies with their implications:

\begin{enumerate}[label=(\alph*)]
    \item \textit{Central Process Mining}: Traditional process mining algorithms are based on a single central event log. Dealing with sensor data, those must be transferred to a central instance to perform the process mining algorithms there.
    \item \textit{Decentral Process Mining}: More than a single central node exists in the decentral paradigm. Data within an organizational unit are aggregated at a central point. Several central nodes must cooperate to create the whole process model. 
    \item \textit{Distributed Process Mining}: In Distributed Process Mining no central entities exist. The data is spread among the data sources. When data from other data sources is needed, they must be queried by a network request. The combined process model may be retrieved at any edge node.  
\end{enumerate}

\begin{figure}[H]
\centering
\begin{subfigure}{.33\textwidth}
    \centering
    \includegraphics[width=\textwidth]{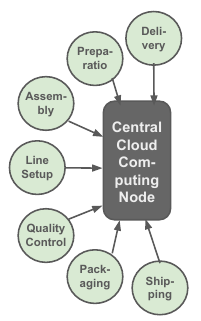}
    \caption{Centralized PM}
\end{subfigure}%
\begin{subfigure}{.33\textwidth}
    \centering
    \includegraphics[width=\textwidth]{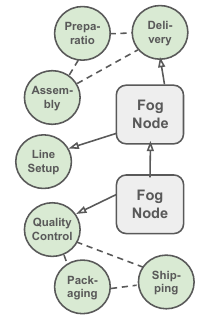}
    \caption{Decentralized PM.}
\end{subfigure}%
\begin{subfigure}{.33\textwidth}
    \centering
    \includegraphics[width=\textwidth]{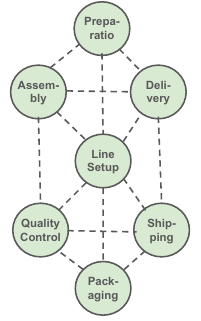}
    \caption{Distributed PM}
\end{subfigure}%
\caption[short]{The three process mining topologies differ in computing and communication allocation, with Distributed PM leveraging data sources' inherent resources.}
\label{fig:topologies}
\end{figure}

\subsection{Extending the Stream Process Mining Model to the CPS domain}
Our proposed model for DPM in cyber-physical systems builds upon the event stream model. To accommodate the unique characteristics of distributed CPS, three key extensions are necessary:

\begin{enumerate}
    \item \textbf{Distributed Data Stream}: Instead of a centrally available event stream, each source node emits a localized event stream. This notion captures the distributed nature of data within CPS.
    \item \textbf{Computing Resource Modeling}: While stream process mining has addressed computational constraints like runtime and memory limitations, these constraints have not been rigorously modeled. Moreover, in the context of CPS, the specific hardware on which computations are performed significantly impacts computing quality measures. 
    \item \textbf{Communication Network}: Devices in CPS communicate via networks. This network communication influences algorithm execution time and must be considered alongside the characteristics in computational resources.
\end{enumerate}
Based on Requirement 1, we formulate definitions for a distributed event stream:
\begin{definition} [Distributed Event Stream]
Let $l \in L$ be a location. The event stream $S_l$ located at $l$ is defined as a subset of the event stream $S$, where all events are located at $l$. $S_l := \{e \in S(i) | location(e) = l\}.$ We refer to the union of all Located Event Streams as a Distributed Data Stream: $S_D := \bigcup_{l \in L} S_l$. 
\end{definition}
We model a computing node as an abstraction comprising a processor, memory, and a network interface. These components can vary significantly regarding their capabilities and performance characteristics. To capture this variability, we introduce a cost model where operations are associated with costs that can differ across devices. Our model avoids using physical units and instead relies on a virtual cost function. This choice is motivated by our primary objective, not to obtain precise performance metrics but to establish a relative performance measure for different devices, thus enabling qualitative comparisons between algorithms and topologies.

Costs are partitioned into two constituent elements: a temporal component and a utilization component. The temporal component quantifies the time required for the operation to complete, whereas the utilization component describes the ratio of the resource's capacity consumed by the operation. To illustrate, storing a 1GB file could require one second of processing time and result in a 10\% utilization of a 10GB hard drive. 

To enable interaction with hardware components within the algorithm, we propose using Hardware Interaction Instructions (HIIs). These instructions serve as the foundation for deriving quality metrics. HIIs' associated costs encompass temporal and utilization components in their computing, storing, and sending operations. The compute operation represents interactions with the processor, the store operation pertains to interactions with memory, and the send operation relates to interactions with the network. Additionally, we incorporate a payload factor to account for the varying complexities of different interactions. The payload reflects the computational and resource demands associated with each interaction. For example, transmitting a large file requires greater network bandwidth than transmitting a smaller one.
\begin{definition}[Hardware Interaction Instruction]
We define the set of Hardware Interaction Instructions as $\mbox{HII}=\{compute, store, send\}$ with
$\forall f \in \mbox{HII}: \mathbb{R}^2 \rightarrow \mathbb{R}^2$ defined as $ f(p,q) = (p \cdot time, q \cdot util)$ where $time > 0$ and $0 \leq util \leq 1$.
\end{definition}
A node is formally defined as a collection of HIIs. Notably, a node can be associated with multiple HIIs of the same kind. This feature proves essential when considering network access scenarios involving connections to diverse participants via distinct network links. Moreover, this approach differentiates between storage types, such as in-memory or disk-based storage.
\begin{definition}[Computing Node]
A Computing Node ($CN$) is defined as a set of HIIs: $CN \subseteq \mbox{HII}^*$.
\end{definition}
\section{\name}

This section presents an evaluation approach for DPM algorithms, which serves as the base of \name. \name accepts as input a DPM algorithm and a corresponding topology provided by an algorithm engineer. The output comprises a set of predefined quality metrics. In the context of this algorithm, all hardware interactions are represented using a unified modeling construct known as an HII. It is crucial to note that only those instructions that significantly contribute to the computational complexity of a problem are exchanged using HIIs. Control flow structures and transient auxiliary variables should be excluded from this modeling approach.

For the demonstration, we introduce the baseline distributed process discovery algorithm. Further, we evaluate it on a setup with three computing nodes. The \textit{baseline edge process discovery algorithm} is stated in Algorithm 1. Whenever it receives an event, it will be stored in a local event log. In the next step, all network-reachable nodes are requested to send the latest recorded event with a given case ID. All potentially preceding events are collected and the latest event is computed. 
The direct-follow relation is derived from the activity of the latest and the incoming event. Finally, the directly follows relation is stored.

\begin{algorithm}[h]
\caption{Baseline Distributed Process Discovery DFG}\label{alg:cap}
\begin{algorithmic}
\Procedure{receive}{event}
 \State $store_{eventLog}(event)$
 \State precedingEvents = list()
 \For{node $\in$ network}
    \State $precedingEvents.append(send_{requestLastEventWithCaseId}^{node}( case(event)))$
 \EndFor
 \State $lastActivity \gets compute_{latestTimestamp}(precedingEvents)$
 \State $dfr \gets (activity(lastActivity), activity(event))$
 \State $store_{followsRelation}(dfr)$
\EndProcedure

\Procedure{requestLastEventWithCaseId}{caseId}
 \State \Return $store_{getLatestEventWithCaseId}(caseId)$ 
\EndProcedure

\Procedure{request}{model}
 \State $model \gets store_{getFollowsRelations}$
\EndProcedure
\end{algorithmic}
\end{algorithm}

To evaluate the algorithm, computing nodes need to be specified and the payload has to be assigned to the corresponding methods.
For this demonstration, we assume three computing nodes $CN_i, i \in \{0,1,2\}$:\\
\begin{align}
CN_i = \{compute(p,q)=(2p,\frac{q}{100}), store(p,q)=(3p,\frac{q}{100}), send^x(p,q)=(10p,\frac{q}{20})\}\nonumber
\end{align}

$payload(m, args) = \begin{cases}
(len(args), len(args)), \text{if m=\textit{latestTimestamp}}\\
(1,1), \text{if m=\textit{eventLog}}\\
(1,2), \text{if m=\textit{followsRelation}}\\
(1,0), \text{if m=\textit{getLatestEventWithCaseId}}\\
(1,0), \text{if m=\textit{getFollowsRelation}}\\
(1,1), \text{if m=\textit{requestLastEventWithCaseId}}\\
\end{cases}$ 

\bigskip

\noindent
The distributed trace resulting from the algorithm execution is as follows:
\begin{enumerate}[itemsep=2mm]
    \item $CN_0$: $store_{eventLog}(e=(c_0, a_1, t_1, n_0)$)
    \item $CN_0$: $send^{\ CN_1}_{latestEventWithCase}(c_0$)
    \item $CN_1$: $store_{getLatestEventWithCaseId}(c_0$)
    \item $CN_0$: $send^{\ CN_2}_{latestEventWithCase}(c_0$)
    \item $CN_2$: $store_{getLatestEventWithCaseId}(c_0$)
    \item $CN_0$: $compute_{latestTimestamp}([(c_0, a_0, t_0, n_1)$])
    \item $CN_0$: $store_{followsRelation}((a_0, a_1))$)
\end{enumerate}

The evaluation of receiving event $e=(c_0, a_1, t_1, n_0)$ by $CN_0$ is as follows. 
We demonstrate our evaluation methodology by determining the temporal cost of processing event e.\\

\begin{align}
\label{eq-time-example}
time(\mathcal{A}, e) \nonumber&= \sum_{step \in exec(\mathcal{A}, e)} time(step) \\ \nonumber
&=time(store(payload(\textit{eventLog},(e=(c_0, a_1, t_1, n_0)))))\\ \nonumber
&+time(send^{CN_1}(payload(\textit{latestEventWithCase},(c_0))))\\ \nonumber
&+time(store(payload(\textit{getLatestEventWithCaseId},(c_0))))\\ \nonumber
&+time(send^{CN_2}(payload(\textit{latestEventWithCase},(c_0))))\\ \nonumber
&+time(store(payload(\textit{getLatestEventWithCaseId},(c_0))))\\ \nonumber
&+time(compute(payload(\textit{latestTimestamp},([(c_0, a_0, t_0, n_1)]))))\\ \nonumber
&+time(store(payload(\textit{followsRelation},((a_0, a_1)))))\\ \nonumber
&=3+10+3+10+2+3\\ \nonumber
&=31
\end{align}

\subsection{Edge computing quality measures}
In the following we model a metric for each instance of those quality attributes.  The process mining-specific metrics of replay fitness, precision, generalization, simplicity or and F1-score already \cite{Foundation} cover the functional correctness. For the remaining metrics, the processing time, resource utilization and scalability in terms of load capacity and resource demand are introduced.

We define processing time as the elapsed time from initiating an event until its completion. This metric is computed by accumulating the execution times of all HIIs that contribute to the processing of the event within a given trace of the algorithm. We further incorporate the time necessary to transmit the outcome to the end-user to obtain the overall response time.
\begin{definition}[Processing Time]
The processing time for event $e$ is defined as: $t_{p} = \sum_{step\in exec(A,e)}t(step)$
\end{definition}
%
Resource utilization is measured individually for each computing resource, CPU, memory, and network. It is quantified by summing the resource utilization of each step in the algorithm and multiplying it by the duration between two consecutive events. Hence, resource utilization increases if more events must be processed in a shorter time range. Memory represents an exception, as it is time-independent. It is occupied until the data is explicitly deleted. This behaviour contrasts with the CPU, whose processing capacity scales (normally) linearly with time.
\begin{definition}[Resource Utilization]
The resource utilization of the computing resource $c$ cpu and network is defined as: $r_c(e) = \frac{1}{\Delta t} \cdot {\sum_{step\in exec(\mathcal{A},e)}r_c(step)}$. The memory utilization is described as: $r_c(e) = {\sum_{step\in exec(\mathcal{A},e)}r_c(step)}$
\end{definition}
Based on the identified quality attributes, the algorithm's scalability on the given computing topology can be derived. In performance engineering, the scalability of a system is evaluated via a Service Level Objective (SLO)~\cite{Henning2022}. An SLO is a function that verifies that a quality measure is constantly fulfilled within a given observation period. An example of an SLO is that the resource utilization of a component does not exceed 95\%. Scalability is characterized by two dimensions: \textit{load capacity} and \textit{resource demand}. Load capacity assesses the maximum sustainable load, ensuring the SLO is still met. Hence, it answers the question of how many requests can be handled by the current infrastructure. On the other hand, resource demand assesses the minimal hardware resource requirements to deal with a given load profile. The scalability determines the relationship between hardware resources and load profile. Thus, it measures the effect of increasing the load at a given point on the resource demand and vice versa.
\begin{definition}[Scalability]
An SLO is a function $slo: L\times CN \rightarrow \{true,false\}$.
The resource demand is defined as $demand(l) = min\{cn \in CN\text{ | }slo(l, cr) = true\}$ while the load capacity is $capacity(cn) = max\{l \in L\text{ | }slo(l, cn) = true\}$.
\end{definition}
\section{Evaluation}
To illustrate the capabilities of our formalism, we conduct a comparative analysis of we examine the three topologies introduced in Figure~\ref{fig:topologies} and an algorithmic optimization inspired by Andersen et al.~\cite{Andersen}. To perform the evaluation, we developed the \name tool, which is publicly available on Github\footnote{\url{https://github.com/HenryWedge/DistributedEnvironmentBuilder/tree/caise}}. The data for the evaluation is generated with the \textit{Distributed Event Factory}~\cite{reiter2024}, an event data generator specialized in distributed event streams. 

\subsection{Benchmark Set up}
A comprehensive description of the benchmark we introduce the algorithms in further detail:

\begin{enumerate}[label=(\alph*)]
    \item \textbf{DFG-miner cloud.} (\textit{yellow} in Figures~\ref{fig:utilization},~\ref{fig:rdlc}), as shown in Figure~\ref{fig:topologies}a. Whenever an event is received by an edge node it is send to the central cloud node. The cloud node updates its directly follows graph on every incoming event.
    \item \textbf{DFG-miner fog.} (\textit{blue}), as shown in Figure~\ref{fig:topologies}b. When an event is received by an edge node they communicate within their subnet to build a partial directly follows graph. When the DFG is requested by the user, the fog node computes the combined DFG of its subnet and requests the other fog nodes for the DFGs of their subnet. Finally, all subnet DFGs are merged to a single one and are given to the user.
    \item \textbf{DFG-miner edge.} (\textit{magenta}), as shown in Figure~\ref{fig:topologies}c the sensors communicate without central instance. On an incoming event, an edge node tries to find its predecessor among all other edge nodes and builds a partial DFG based on that information. When the overall DFG is requested the requested edge node collects all partial DFGs from all edge nodes and merges them together. The implementation follows Algorithm~\ref{alg:cap}.
    \item \textbf{EdgeMiner.} (\textit{red}), also follows the topology of Figure~\ref{fig:topologies}c. It implements a simplified version of the EdgeMiner by Andersen et al.~\cite{Andersen} It optimizes the \textit{DFG-miner edge} by requesting the edge node which has been the predecessor within the most cases. When it does not serve an event of the requested case, the second most predecessor is retrieved and so on.  
\end{enumerate}

We use three different types of computing nodes for edge, fog and cloud. The fog node has 50\% and the cloud node 100\% more CPU and memory capabilities compared to the edge node. The connection between an edge node and the cloud nodes takes 5 times longer than the connection between the edge node. Transferring between the edge node and fog node takes 1.5 times as long. For every ten emitted events the DFG is requested once within a benchmark run.

\subsection{Benchmark Results}

\begin{figure}[h]
\centering
\begin{subfigure}{.5\textwidth}
    \centering
    \includegraphics[width=\textwidth]{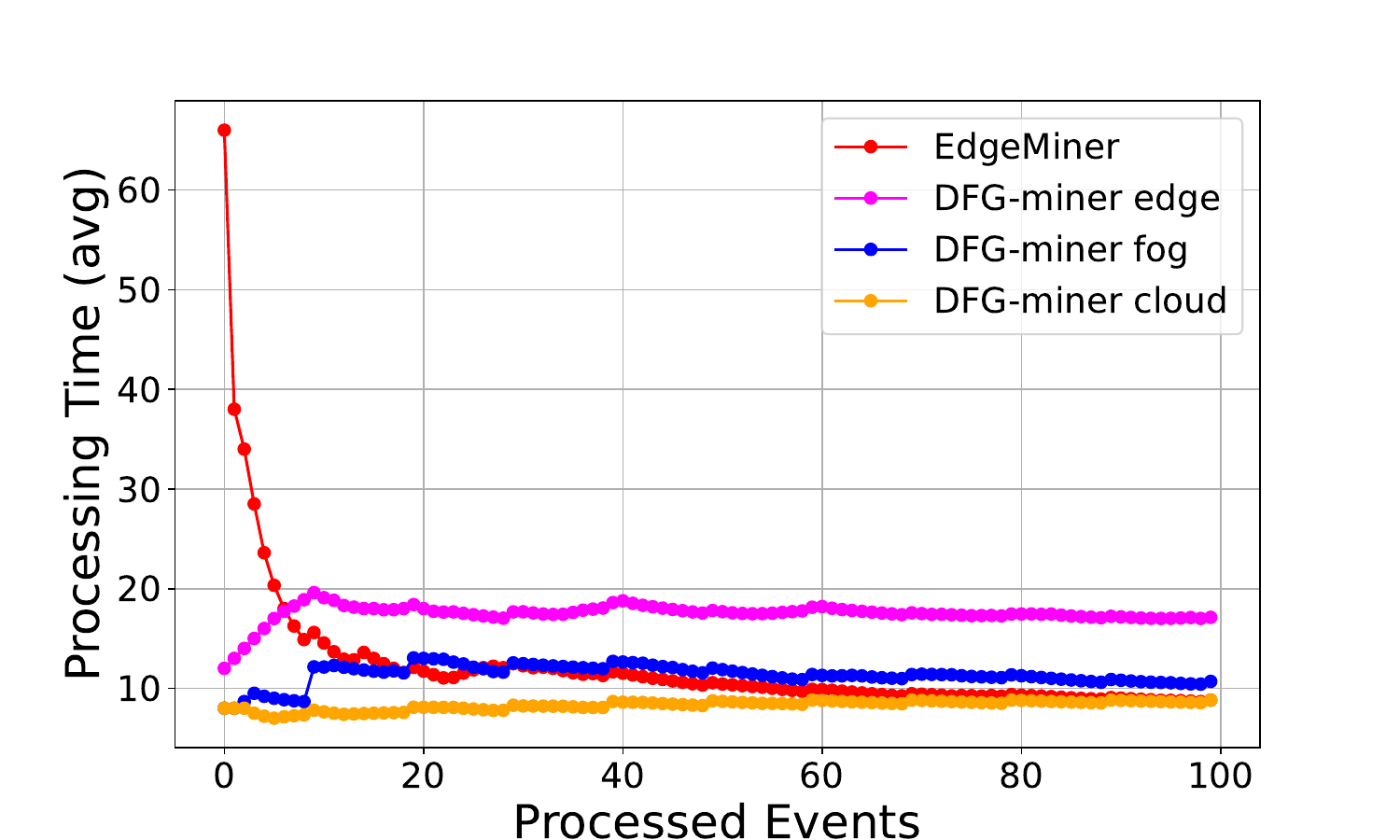}
    \caption{Average processing time}
\end{subfigure}%
\begin{subfigure}{.5\textwidth}
    \centering
    \includegraphics[width=\textwidth]{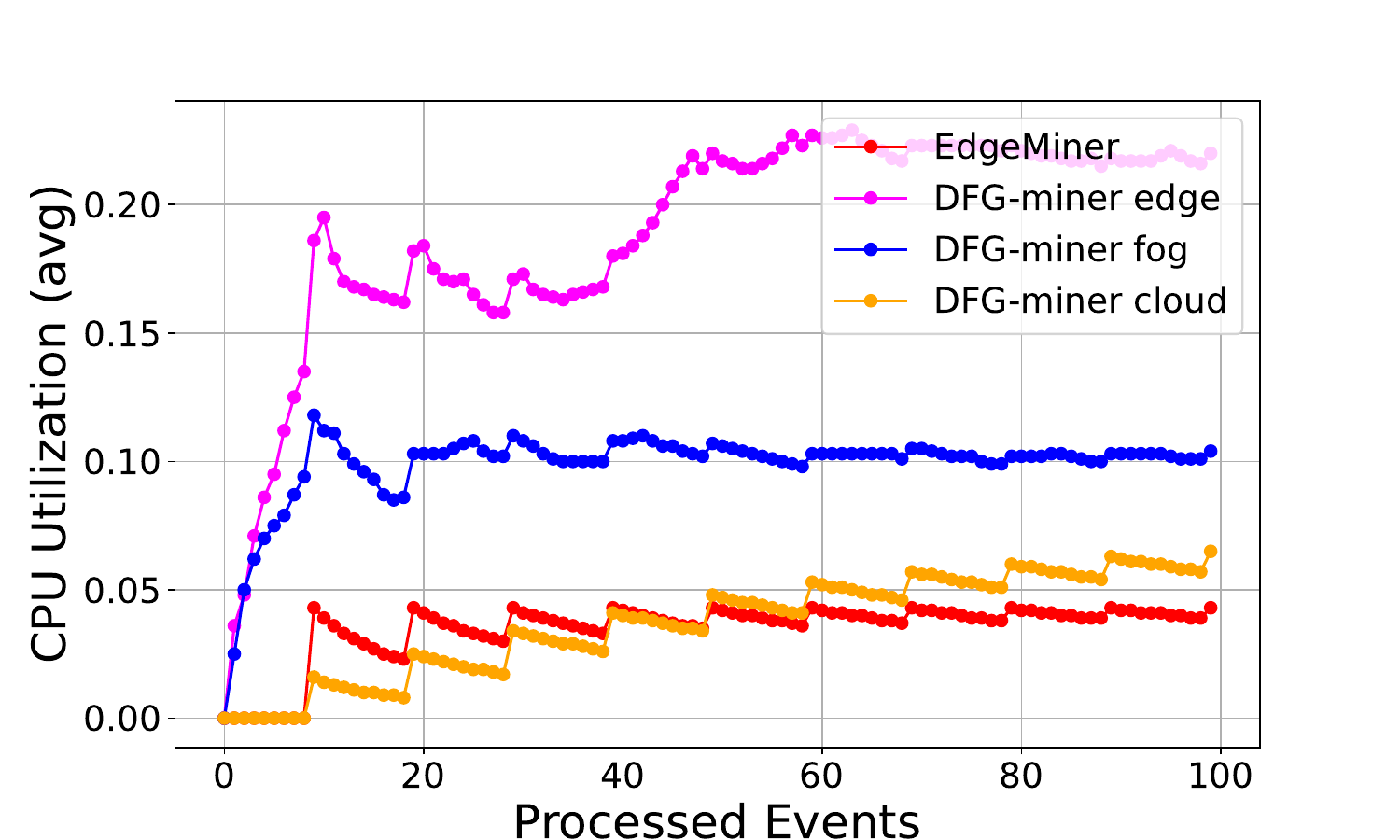}
    \caption{Average CPU utilization}
\end{subfigure}
\begin{subfigure}{.5\textwidth}
    \centering
    \includegraphics[width=\textwidth]{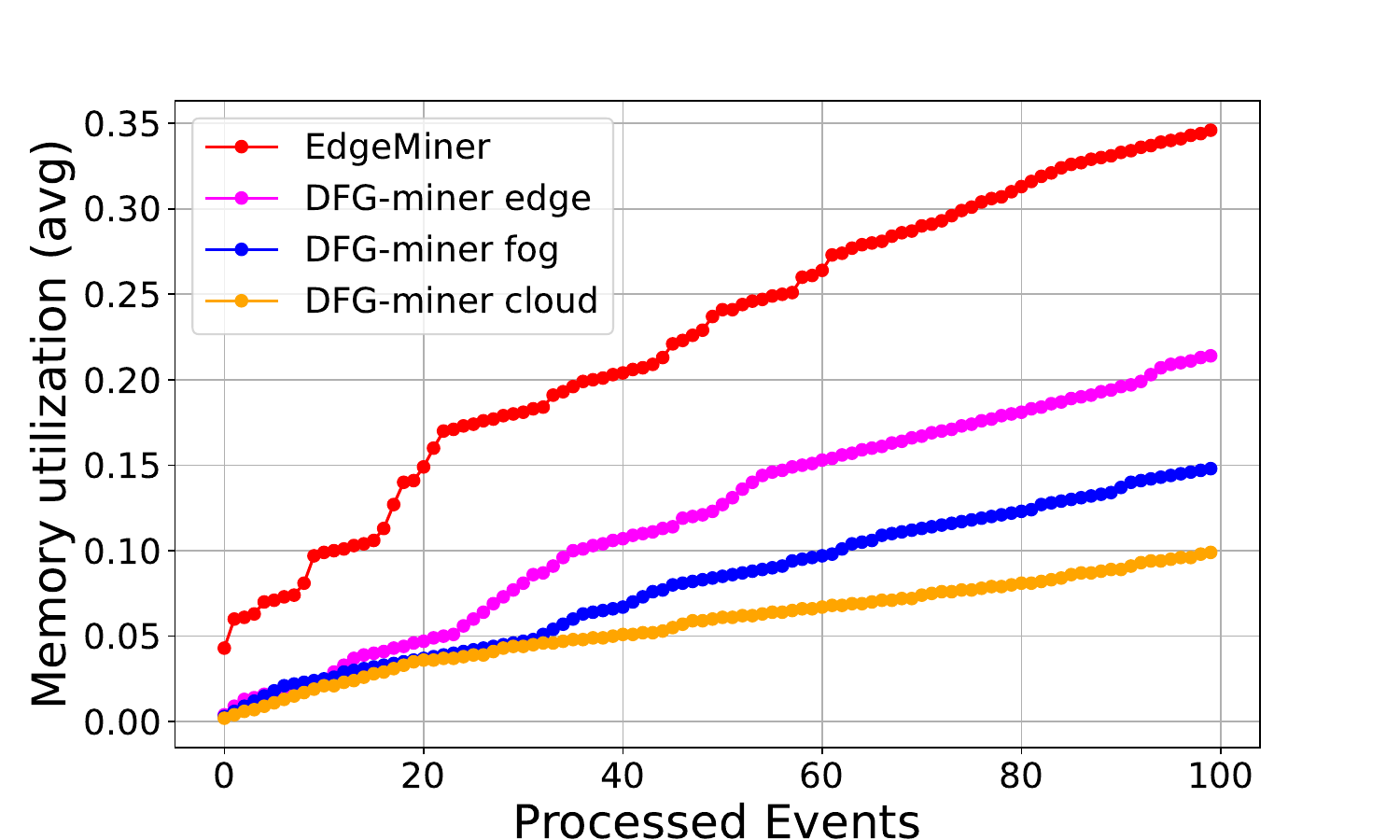}
    \caption{Average memory utilization}
\end{subfigure}%
\begin{subfigure}{.5\textwidth}
    \centering
    \includegraphics[width=\textwidth]{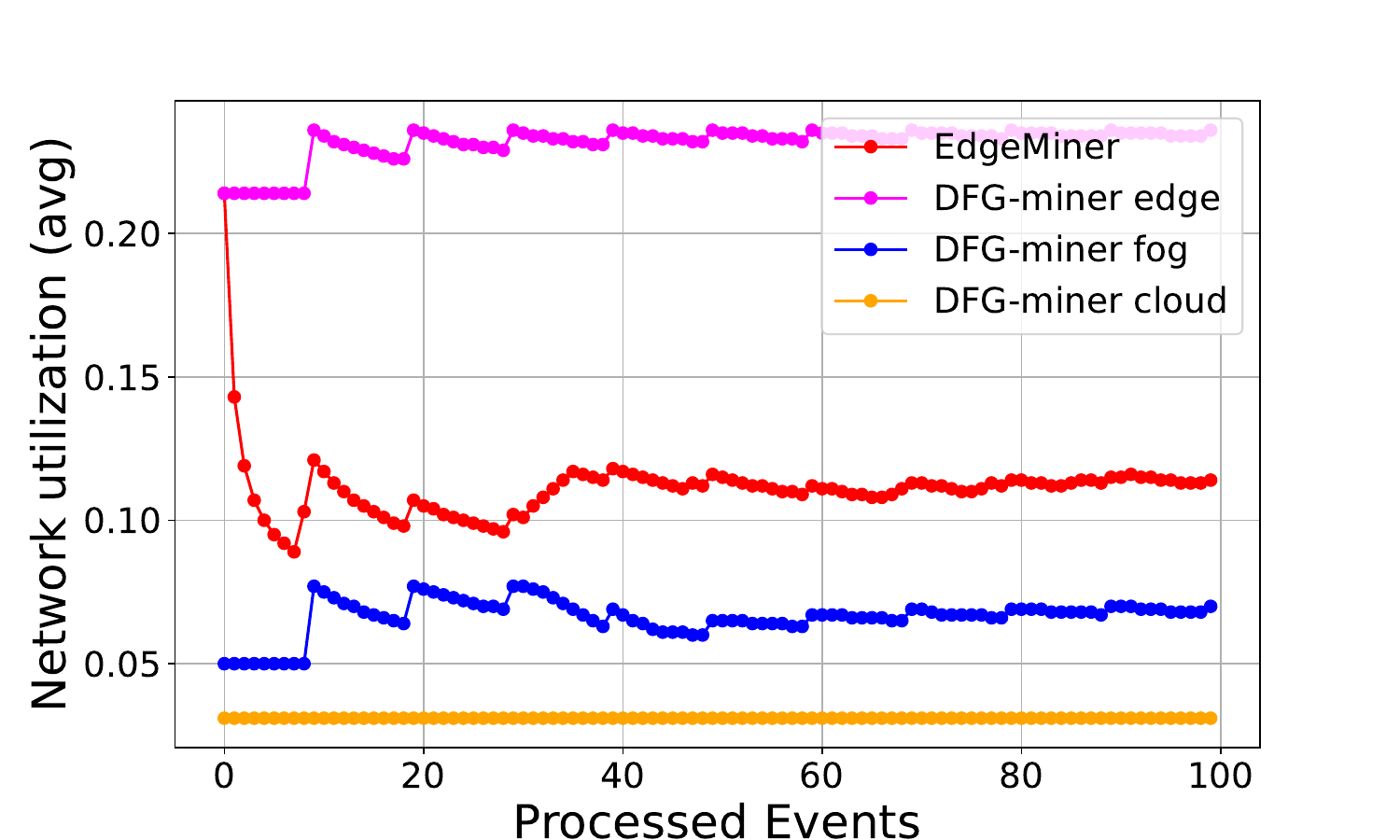}
    \caption{Average network utilization}
\end{subfigure}
\caption[short]{Quality measures derived from \name by processed events. It captures significant changes between algorithms and topologies.}
\label{fig:utilization}
\end{figure}

The results of the benchmark are shown in Figure~\ref{fig:utilization}. There the average processing time as well the average utilization for CPU, memory and network are shown.

Network utilization is highest at the DFG-miner edge. It remains constant during event reception and experiences a significant spike when the process model is queried. This is attributable to the transmission of all partial DFGs over the network. The diminishing height of the spike is a consequence of averaging. In contrast, the EdgeMiner succeeds in minimizing network costs. Although the EdgeMiner initially exhibits high utilization due to querying all edge nodes, this rapidly decreases over time. Minor upward spikes correspond to the initiation of new cases, requiring fresh queries to all edge nodes. The Fog topology further reduces network costs, displaying a pattern similar to the DFG-miner edge in the graph trajectory. The DFG-miner cloud exhibits constant network utilization, as each event is sent directly to the cloud node and the DFG is already present locally on the node. Consequently, querying the DFG does not necessitate additional network requests.

Regarding memory utilization, a continuous increase in values is observed across all variants. This is attributed to the absence of active data deletion. Consequently, a point may be reached where memory demand exceeds capacity. Notably, the EdgeMiner exhibits higher memory consumption due to the additional storage of metadata such as frequent predecessors. 

CPU utilization reveals two distinct clusters. The DFG-miner edge and fog must continually compare timestamps for each event to compute direct follow relationships, resulting in consistent CPU demands. In contrast, the other two variants experience CPU-intensive computations primarily during DFG generation.

The average processing time accumulates interactions with hardware components. While the EdgeMiner initially exhibits the highest processing time, it subsequently improves to achieve the lowest processing time. This demonstrates that algorithmic optimizations can enable DPM to compete with centralized process mining at the computational level. Conversely, unoptimized approaches like the DFG-miner edge exhibit the longest processing time.

\begin{figure}[h]
\centering
\begin{subfigure}{.5\textwidth}
    \centering
    \includegraphics[width=\textwidth]{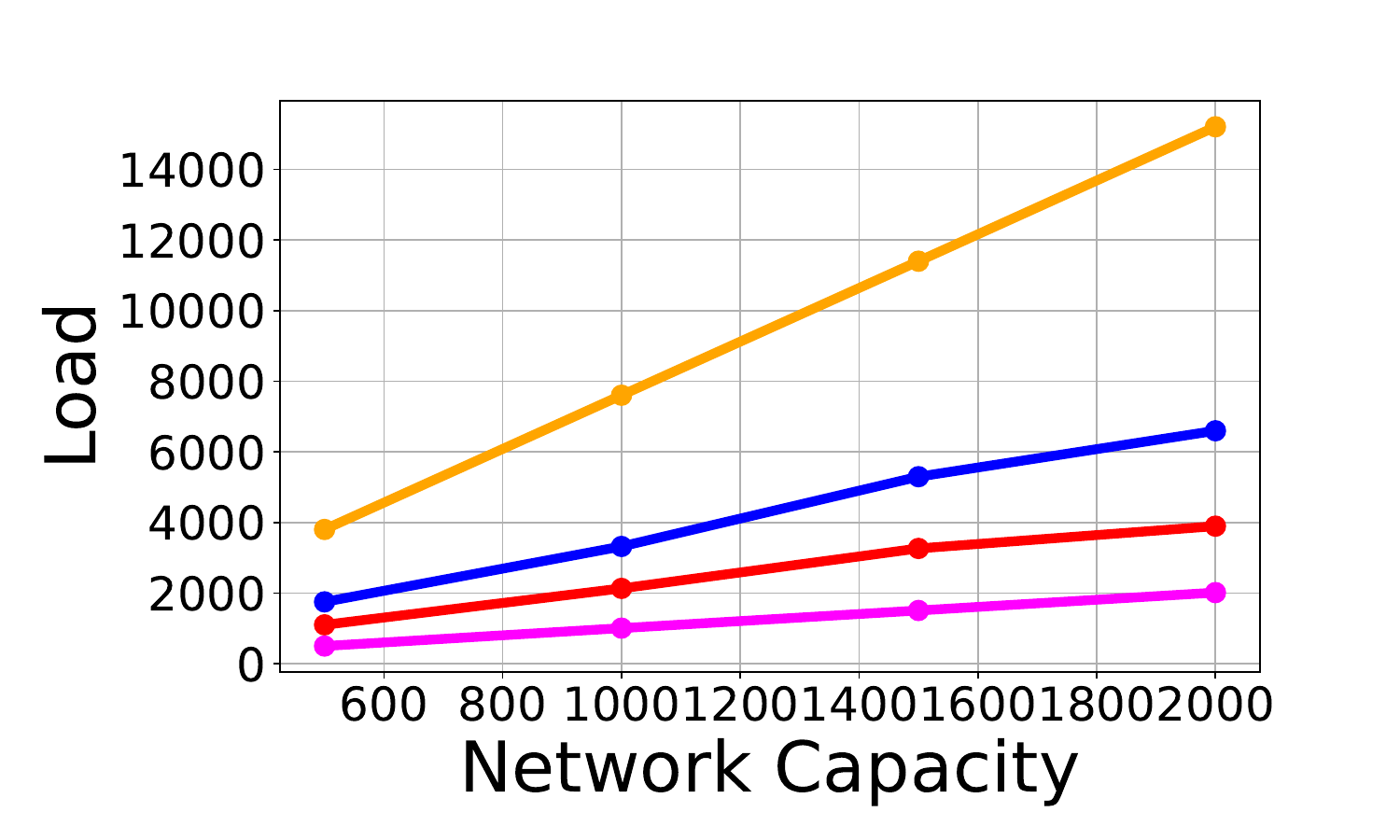}
    \caption{Load Capacity}
\end{subfigure}%
\begin{subfigure}{.5\textwidth}
    \centering
    \includegraphics[width=\textwidth]{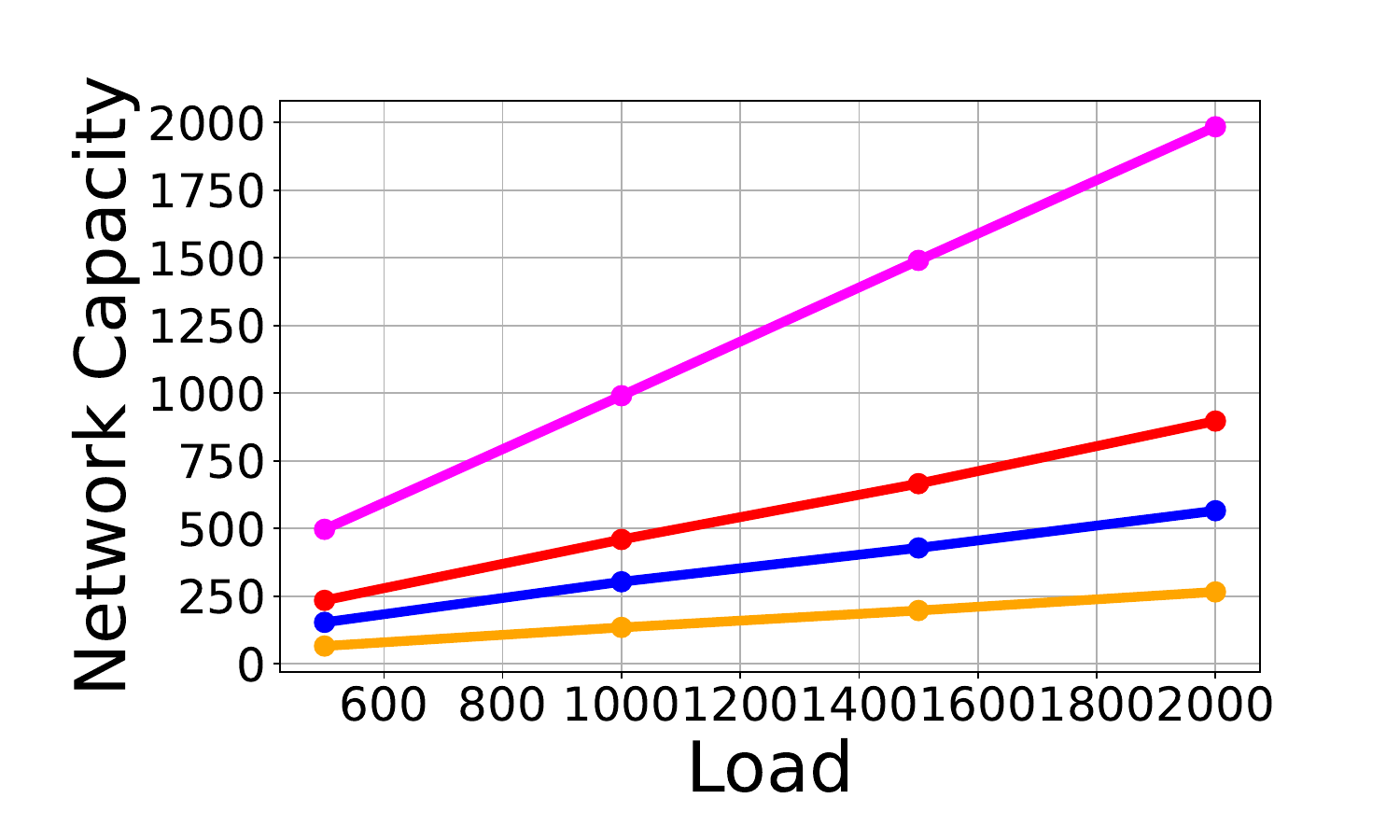}
    \caption{Resource Demand}
\end{subfigure}
\caption[short]{\label{fig:rdlc}Resource Demand and Load Capacity in relation to the provisioned network capacity}
\end{figure}

For the scalability benchmark presented in Figure~\ref{fig:rdlc}, the metrics load capacity and resource demand were determined. The SLO ensures that network utilization must not exceed 95\%. The results indicate that, in terms of network utilization, the cloud continues to exhibit the best scalability. An information systems engineer seeking to evaluate the need for additional hardware for distributed process mining or to determine the maximum load for an existing setup can consult a visualization equivalent to Figure~\ref{fig:rdlc}.

\section{Discussion}
The objective of \name is been to provide a qualitative comparison of the computational aspects among different algorithms and topologies.
For a deeper understanding of the behavior of concrete algorithms we recommend to run \name with various data that and not solely on generated data. Hence, the representational bias to the specific dataset is reduced. Further, classical process mining quality metrics such as recall, fitness, generalization, and simplicity should be determined in order to evaluate newly developed distributed process mining algorithms. Furthermore, the current version of \name primarily uses average values for evaluating processing times and resource utilization. We argue that \name aims to provide a simple mathematical abstraction for use in algorithm engineering. However, this simplicity necessitates the omission of specific details. Details omitted for simplicity include dependencies between processing times and resource utilization, routing between sensors, asynchronous communication, and hardware faults. For more realistic evaluations, it is planned to extend \name to enable evaluations on actual hardware.

\section{Conclusion}

In this paper, we introduced a formal model and a benchmark for distributed process mining and demonstrated its relevance in the domain of cyber-physical systems. We extended existing formalisms for DPM and the streaming model to incorporate data locality, computing resources, and network communication. Based on these extensions, we developed \name, a benchmark for comparing DPM algorithms. Our evaluation results show that \name is suitable for comparing DPM algorithms and topologies against each other. Furthermore, it provides insights into algorithmic behavior. This lays the foundation for the development and evaluation of future DPM algorithms. 

Our future work will focus on enhancing the algorithm and topology evaluations within the \name framework. To this end, we plan to address the limitations identified in the discussion section and extend the simulation capabilities to include real hardware execution. Moreover, we will explore additional motivations for Distributed Process Mining, such as privacy and minimization of network-transferred data.

\begin{credits}
\subsubsection{\ackname} This work received funding from the Deutsche Forschungsgemeinschaft (DFG), grant 496119880
\end{credits}
\newpage
%
%
%
\bibliographystyle{splncs04}
\bibliography{bibliography}